\documentclass[cameraready]{Interspeech}
\title{Iterate to Differentiate: Enhancing Discriminability and Reliability in Zero-Shot TTS Evaluation}

\author[affiliation={1,2}]{Shengfan}{Shen}
\author[affiliation={2}]{Di}{Wu}
\author[affiliation={2}]{Xingchen}{Song}
\author[affiliation={2}]{Dinghao}{Zhou}
\author[affiliation={3}]{Liumeng}{Xue}
\author[affiliation={2}]{Meng}{Meng}
\author[affiliation={2}]{Jian}{Luan}
\author[affiliation={1}, correspondingauthor]{Shuai}{Wang}
\address{
    $^1$ Nanjing University, China \\
    $^2$ MiLM Plus, Xiaomi Inc., China \\
    $^3$ Hong Kong University of Science and Technology, China
}

\email{shenshengfan@hnu.edu.cn,shuaiwang@nju.edu.cn}
\keywords{speech synthesis, iterative evaluation, human-aligned metrics}

\usepackage{comment}
\usepackage{multirow}
\usepackage[table]{xcolor}
\usepackage{booktabs}
\usepackage[normalem]{ulem}
\usepackage[most]{tcolorbox}
\usepackage{algorithm}
\usepackage{algpseudocode}
\usepackage{float}
\begin{document}

\maketitle

% the abstract here must exactly match the abstract entered into the paper submission system
\begin{abstract}
    % 1000 characters. ASCII characters only. No citations.
    
    Reliable evaluation of modern zero-shot text-to-speech (TTS) models remains challenging. Subjective tests are costly and hard to reproduce, while objective metrics often saturate, failing to distinguish SOTA systems. To address this, we propose Iterate to Differentiate (I2D), an evaluation framework that recursively synthesizes speech using the model's own outputs as references. Higher-quality models exhibit greater resilience to the distributional shift induced by iterative synthesis, resulting in slower performance degradation. I2D exploits this differential degradation to amplify performance gaps and reveal robustness. By aggregating objective metrics across iterations, I2D improves discriminability and alignment with human judgments, increasing system-level SRCC from 0.118 to 0.464 for UTMOSv2. Experiments on 11 models across Chinese, English, and emotion datasets demonstrate that I2D enables more reliable automated evaluation for zero-shot TTS.

\end{abstract}

\section{Introduction}

In recent years, text-to-speech (TTS) has made significant progress \cite{anastassiou2024seed,du2025cosyvoice,zhou2025indextts2,hu2026qwen3}, largely driven by advances in generative modeling, such as large language models (LLMs) and diffusion models, as well as the rapid expansion of training data and computational resources. Modern TTS systems, particularly in zero-shot voice cloning scenarios, are now capable of producing highly natural and expressive speech that is often indistinguishable from human speech.

Despite these advances in TTS modeling, evaluation methodologies have not kept pace with improvements in synthesis quality \cite{yang2025towards}. Traditional TTS evaluation can be broadly categorized into objective and subjective approaches. Objective evaluation typically relies on metrics such as word error rate (WER) and speaker similarity (SIM). However, for state-of-the-art (SOTA) TTS systems, these metrics are increasingly prone to saturation, where marginal improvements in objective scores often fail to translate into perceptible gains in human perception \cite{tee2025sp}. This issue is exacerbated by the inherent errors and biases of the evaluation models themselves, which limit their ability to reliably distinguish between high-quality synthetic samples. Subjective evaluation, by contrast, assesses audio quality through listening tests, most commonly using the Mean Opinion Score (MOS), where listeners rate speech quality on a five-point scale. While MOS generally reflects human preferences reasonably well, its subjectivity and inter-rater variability make the results difficult to reproduce \cite{chiang2023we}. Furthermore, the substantial time and financial costs associated with human evaluation severely restrict its scalability. To alleviate these issues, recent studies have explored neural network–based MOS prediction models as surrogates for human ratings \cite{lo2019mosnet,reddy2021dnsmos,saeki2022utmos}. Nevertheless, these methods still exhibit weak correlation with human judgments, particularly on out-of-domain data or when distinguishing subtle quality differences \cite{huang2024voicemos,wang2026urgentmos}.

To address these challenges, we propose \textbf{Iterate to Differentiate (I2D)}, a novel evaluation framework that transcends static assessment. This strategy leverages the zero-shot capabilities of modern TTS models by recursively using synthesized outputs as reference audio and re-synthesizing them over multiple iterations, thereby exploiting error accumulation effects to progressively amplify performance differences across models. By analyzing the evolution of objective metrics across multiple synthesis stages, I2D effectively characterizes a model's underlying robustness and quality. Our findings reveal that the aggregated performance trajectory over iterations provides a more reliable human-aligned proxy than single-turn objective scores. Our contributions can be summarized as follows:

\begin{itemize}

\item We conduct a systematic analysis of existing objective evaluation metrics and demonstrate that, under the conventional evaluation protocol, they suffer from severe score saturation across all metrics. This saturation leads to unreliable model rankings, particularly for predictive MOS metrics, limiting their ability to discriminate among SOTA TTS systems.
\item We introduce the I2D framework, which aggregates objective scores across multiple synthesis iterations. This approach amplifies performance differences between models, reduces sensitivity to the inherent errors of evaluation models, and improves correlation with human judgments while reflecting model robustness.
\item We conduct a comprehensive comparative analysis of 11 TTS models across three datasets and provide a detailed examination of performance differences across models.
\end{itemize}

\begin{figure*}[t]
  \centering
  \includegraphics[width=\textwidth]{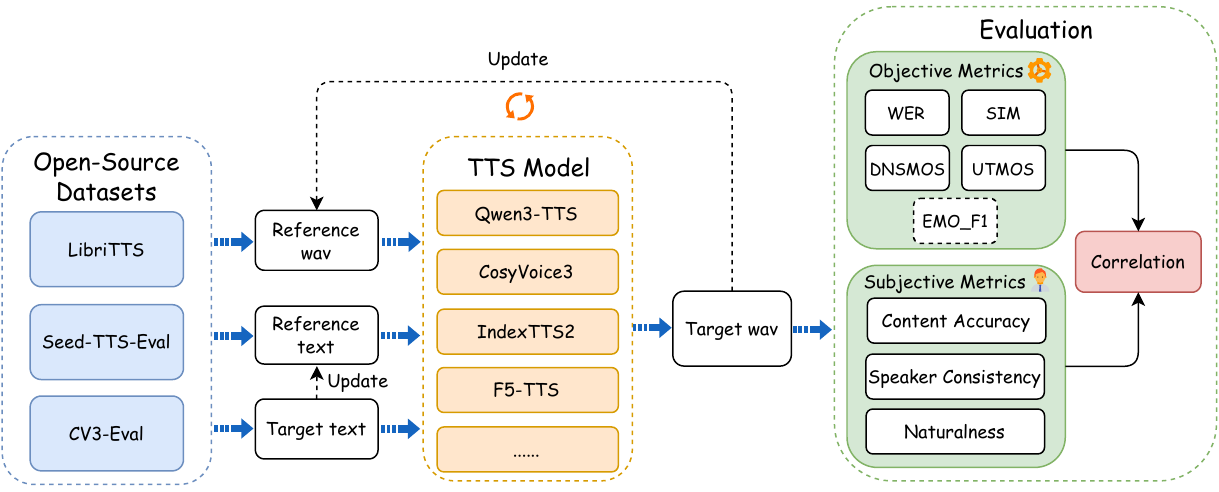}
  \caption{The overall workflow of our evaluation. The dashed arrows indicate that, after the first synthesis, the reference wav and reference text are updated using the target wav and target text. The objective metrics \texttt{EMO\_F1} are applied only to the Emotion dataset.}
  \label{fig:overall}
\end{figure*}

\section{Related Work}
\subsection{Zero-shot TTS Models}

Modern zero-shot TTS systems can be broadly categorized into three architectural paradigms: autoregressive, non-autoregressive, and hybrid models. Since evaluation behavior may vary across architectural families, we include representative systems from each category in our experiments.

Autoregressive (AR) models \cite{hu2026qwen3,chen2024takin,wang2025spark,xie2025fireredtts,liao2024fish,casanova2024xtts} formulate TTS as a conditional sequence generation problem over discrete audio tokens, employing decoder-only Transformers to predict speech tokens step by step. This paradigm tends to yield high naturalness and strong prosodic coherence, at the cost of increased inference latency.

Non-autoregressive (NAR) architectures generate speech representations in parallel, encompassing several sub-paradigms. Diffusion-based models \cite{gao2023e3,ju2024naturalspeech} iteratively refine noisy representations through a learned denoising process. Flow-matching models \cite{chen2025f5} learn continuous normalizing flows via straight-line trajectory regression, achieving competitive quality with fewer sampling steps. Masked generative models \cite{wang2024maskgct} progressively unmask discrete token sequences inspired by masked language modeling.

Hybrid approaches \cite{anastassiou2024seed,du2025cosyvoice,zhou2025indextts2,guo2024fireredtts} adopt a two-stage pipeline in which an LLM first predicts intermediate speech tokens autoregressively, which are subsequently decoded into high-fidelity acoustic features using a diffusion or flow-matching model, effectively combining the respective and complementary strengths of both paradigms.

\subsection{TTS Evaluation Benchmarks and Metrics}

Conventional TTS evaluation relies primarily on Word Error Rate (WER) and Speaker Similarity (SIM), which measure content intelligibility and speaker identity preservation, respectively. Subjective evaluation is most commonly conducted using the Mean Opinion Score (MOS), where listeners rate overall perceptual speech quality on a five-point scale \cite{chiang2023we}. Variants such as CMOS employ pairwise comparisons to capture relative preferences, while SMOS focuses on style and timbre similarity to a target speaker. Although MOS broadly reflects human perceptual preferences, its inherent inter-rater variability and high sensitivity to diverse conditions make scores difficult to reproduce and compare across studies.

To reduce reliance on costly listening tests, prior work has proposed neural predictors that estimate perceptual quality directly from audio signals. DNSMOS \cite{reddy2021dnsmos} introduced a non-intrusive quality predictor originally designed for speech enhancement. The VoiceMOS Challenge 2022 \cite{huang2022voicemos} subsequently advanced this direction and led to widely adopted models such as UTMOS \cite{saeki2022utmos}. The VoiceMOS Challenge 2024 \cite{huang2024voicemos} further introduced a ``zoomed-in'' evaluation subset comprising high-quality systems and revealed that most existing MOS predictors struggle to reliably distinguish and rank strong models. Similarly, UrgentMOS \cite{wang2026urgentmos} identified the same phenomenon, further underscoring the limitations of neural MOS predictors in the strong-model regime.

Inspired by LLM-as-Judge approaches \cite{zheng2023judging}, recent studies have explored Large Speech Language Models (LSLMs) as surrogates for human evaluation. AudioJudge \cite{manakul2025audiojudge} examines the feasibility of unified LSLM-based evaluation and analyzes the effects of prompt engineering. SpeechJudge \cite{zhang2025speechjudge} constructs a large-scale human preference dataset and trains a dedicated judge via a two-stage procedure on Qwen2.5-Omni-7B \cite{xu2025qwen2}. SpeechLLM-as-Judges \cite{wang2025speechllm} further explores structured and explainable quality assessment using LSLMs. While promising, these methods introduce substantial computational overhead and may inherit the biases of the underlying LSLM.

Overall, existing evaluation methods face limitations in the SOTA regime: objective metrics suffer from score saturation and sensitivity to evaluation noise, subjective tests lack scalability, neural MOS predictors struggle to reliably distinguish subtle quality differences, and LSLM-based judges, although promising, remain at an exploratory stage.

\section{Methodology}

Our approach is motivated by a key observation: \textit{under the current evaluation paradigm, objective metrics for SOTA TTS systems often exhibit score saturation, with inter-model differences compressed into a narrow range}. When these small differences become comparable to the intrinsic noise of evaluation models, metric fluctuations may fail to reflect true performance gaps, resulting in unreliable rankings and weak human alignment. To overcome this limitation, we adopt an error accumulation strategy. We recursively reuse a model’s own synthesized outputs as reference inputs, inducing progressive distributional shift through iterative generation. Stronger models degrade more slowly, while weaker models deteriorate faster, leading to amplified performance differences across iterations. By exploiting this differential degradation, our framework restores discriminability to existing objective metrics.

Figure~\ref{fig:overall} illustrates the overall workflow of the I2D evaluation framework. We construct the evaluation data based on three open-source datasets: LibriTTS \cite{zen2019libritts}, Seed-TTS-Eval \cite{anastassiou2024seed}, and CV3-Eval \cite{du2025cosyvoice}. Using the proposed iterative evaluation protocol, we conduct a systematic assessment of 11 SOTA TTS models. During evaluation, we compute a set of objective metrics for all synthesized speech samples and perform human subjective evaluations on a selected subset to analyze the correlation between objective measures and human judgments. This section describes the design of the iterative synthesis protocol, the construction of the evaluation datasets, and the metrics used for both objective and subjective evaluation.

\subsection{Iterative Synthesis Protocol}

Given a TTS model \(M\), we define an iterative synthesis protocol over an evaluation set, where each sample is a triplet \((\texttt{ref\_wav}_i, \texttt{ref\_text}_i, \texttt{text}_i)\). Here, \(\texttt{ref\_wav}_i\) denotes the reference speech, \(\texttt{ref\_text}_i\) its transcription, and \(\texttt{text}_i\) the target text to be synthesized.

At each iteration, the speech generated by \(M\) is reused as the reference audio for the next round, while the target text remains unchanged. Repeating this process yields synthesized samples at increasing iteration depths. Notably, if the model introduces intelligibility errors at iteration \(j\) (e.g., hallucinations or omissions), then the synthesized speech becomes inconsistent with the target text used in iteration \(j+1\). This text-audio mismatch serves as an implicit error amplification mechanism: models with weaker robustness accumulate misalignment more rapidly, leading to faster quality degradation and more pronounced performance differences. Formally, the iterative synthesis process is defined as follows:

\begin{algorithm}[H]
\caption{Iterative Speech Generation}
\label{alg:iterative_generation}
\begin{algorithmic}

\State \textbf{Input:} $M$, $\text{ref\_wav}_i$, 
$\text{ref\_text}_i$, $\text{text}_i$, $\mathrm{max\_iteration}$

\State \textbf{Output:} 
$\{\text{iter\_wav}_i^j \mid j = 1, \dots, \mathrm{max\_iteration}\}$

\State $j \gets 1$

\While{$j \le \mathrm{max\_iteration}$}
    \State $\text{iter\_wav}_i^j \gets 
    M(\text{ref\_wav}_i, \text{ref\_text}_i, \text{text}_i)$
    \State $\text{ref\_wav}_i \gets \text{iter\_wav}_i^j$
    \State $\text{ref\_text}_i \gets \text{text}_i$
    \State $j \gets j + 1$
\EndWhile

\end{algorithmic}
\end{algorithm}

The iteration continues until the predefined maximum number of iterations \(\mathrm{max\_iteration}\) is reached. All synthesized speech samples generated at each iteration are retained and used for subsequent objective metric computation and analysis.

\subsection{Dataset Construction}

We construct three evaluation subsets: a Chinese dataset, an English dataset, and an emotion dataset. The details of each subset are summarized as follows:
\begin{itemize}
\item \textbf{Chinese dataset}: This subset originates from the \textit{test-zh} split of Seed-TTS-Eval, comprising speech samples derived from DiDiSpeech \cite{guo2021didispeech}. It contains 2,020 utterances from 1,010 speakers, with each speaker contributing two audios. The duration of each sample ranges from 4 to 12 seconds.

\item \textbf{English dataset}: This subset is constructed from the \textit{test-clean} split of LibriTTS. We filter the original samples by duration, retaining only those between 3 and 15 seconds. The final dataset consists of 2,915 utterances from 38 speakers.

\item \textbf{Emotion dataset}: This subset is sourced from the \textit{Emotion Cloning} split of CV3-Eval, with speech samples from EmoBox \cite{ma2024emobox} and SeCap \cite{xu2024secap}. It includes both Chinese and English speech and covers three emotion categories: happy, sad, and angry. For each language, it includes 50 samples per emotion, resulting in a total of 300 samples.
\end{itemize}

In addition, we randomly select 100 samples from the Chinese dataset to form a human-evaluation subset, ensuring that each selected sample corresponds to a unique speaker.

\subsection{Evaluation Metrics}

We perform objective evaluations on all datasets and further conduct subjective evaluations on the human-evaluation subset. For objective evaluation, we adopt four commonly used metrics for the Chinese and English datasets: word/character error rate (WER/CER), speaker similarity (SIM), DNSMOS \cite{reddy2021dnsmos}, and UTMOSv2 \cite{baba2024t05}. For the emotion dataset, we report only the emotion classification F1-score. For subjective evaluation, we adopt the Mean Opinion Score (MOS) protocol. To enable fine-grained analysis and correlation with objective metrics, we define three evaluation dimensions: \textit{Content Accuracy}, \textit{Speaker Consistency}, and \textit{Overall Naturalness}. For each dimension, detailed scoring criteria and annotation guidelines are provided. We recruit trained annotators and conduct standardized training prior to annotation to ensure a consistent understanding of the evaluation criteria. During annotation, each test sample is presented together with the synthesized audio, the corresponding reference audio, and the target text, enabling informed judgments. To mitigate individual subjectivity, each sample is independently rated by five or six annotators.

\begin{table}[h]
\centering
\caption{Objective evaluation metrics and subjective evaluation dimensions}
\label{tab:metrics}
\begin{tabular}{ll}
\hline
\textbf{Type} & \textbf{Metrics / Dimensions} \\
\hline
\multirow{5}{*}{Objective} 
 & WER/CER \\
 & SIM \\
 & DNSMOS \\
 & UTMOSv2 \\
 & Emotion F1 \\
\hline
\multirow{3}{*}{Subjective} 
 & Content Accuracy \\
 & Speaker Consistency \\
 & Overall Naturalness \\
\hline
\end{tabular}
\end{table}

To comprehensively account for evaluation results across multiple synthesis iterations, rather than relying on a single iteration, we design several aggregation methods to summarize metric trajectories over iterations. Let \(\text{score}_i\) denote the metric value at the \(i\)-th iteration, and let \(N = \text{max\_iteration}\) denote the maximum number of iterations.

\begin{itemize}
\item \textbf{Mean Score} \\
The arithmetic mean of metric values across all iterations:
\begin{align}
    \text{Mean} = \frac{1}{N} \sum_{i=1}^{N} \text{score}_i
\end{align}

\item \textbf{Linearly Weighted Average (LWA)} \\
Later iterations are assigned linearly increasing weights:
\begin{align}
    \text{LWA} = \frac{\sum_{i=1}^{N} i \cdot \text{score}_i}{\sum_{i=1}^{N} i}
\end{align}

\item \textbf{Exponentially Weighted Average (EWA)} \\
Later iterations are assigned exponentially decaying weights:
\begin{align}
    \text{EWA} = \frac{\sum_{i=1}^{N} \alpha^{i} \cdot \text{score}_i}{\sum_{i=1}^{N} \alpha^{i}}, \quad \alpha = 0.9
\end{align}

\item \textbf{Area Under Curve (AUC)} \\
The iteration-wise metric values are treated as a discrete curve, and the area is computed using the trapezoidal rule:
\begin{align}
    \text{AUC} = \sum_{i=1}^{N-1} \frac{\text{score}_i + \text{score}_{i+1}}{2}
\end{align}

\end{itemize}

For correlation analysis, metrics with lower-is-better semantics, such as WER/CER, are converted to higher-is-better form when necessary (e.g., using \(1-\mathrm{CER}\)) to ensure consistent interpretation across metrics.

\begin{figure*}[t]
  \centering
  \includegraphics[width=.85\textwidth]{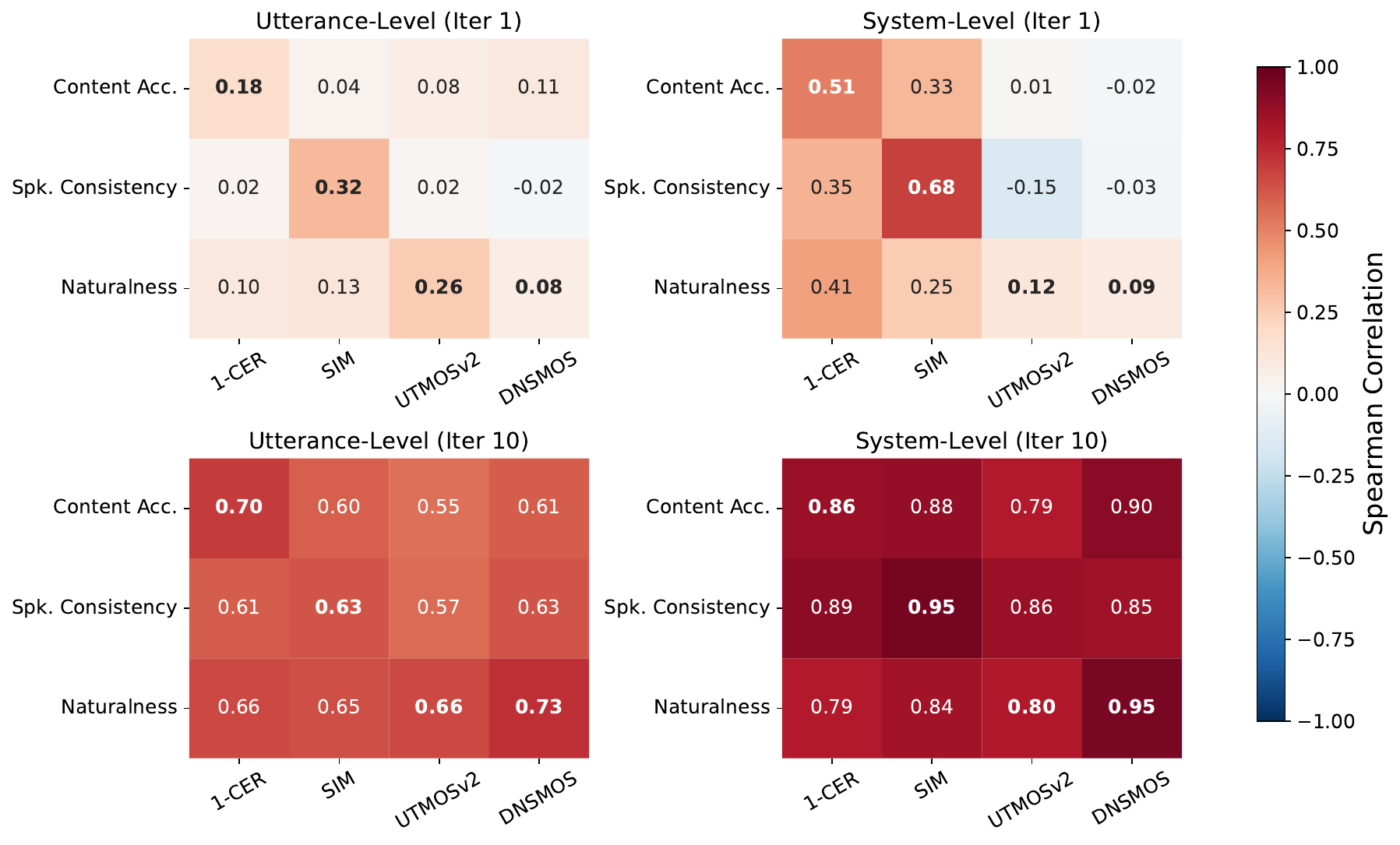}
  \caption{\textbf{SRCC} between objective metrics and subjective dimensions at \textbf{utterance and system levels} (1st and 10th iterations). Utterance-level SRCC is calculated from all individual sample scores, while system-level is derived from model rankings. Specifically, we compare SIM with Spk. Consistency, 1-CER with Content Acc., and UTMOSv2/DNSMOS with Naturalness.}
  % \caption{\textbf{SRCC} between objective metrics and subjective dimensions at \textbf{utterance and system levels} (1st and 10th iterations). Utterance-level SRCC is calculated from all individual sample scores, while system-level is derived from model rankings. Specifically, we compare SIM with Spk. Consistency, 1-CER with Content Acc., and UTMOSv2/DNSMOS with Naturalness.}
  \label{fig:correlation}
\end{figure*}

\begin{figure*}[t]
  \centering
  \includegraphics[width=\textwidth]{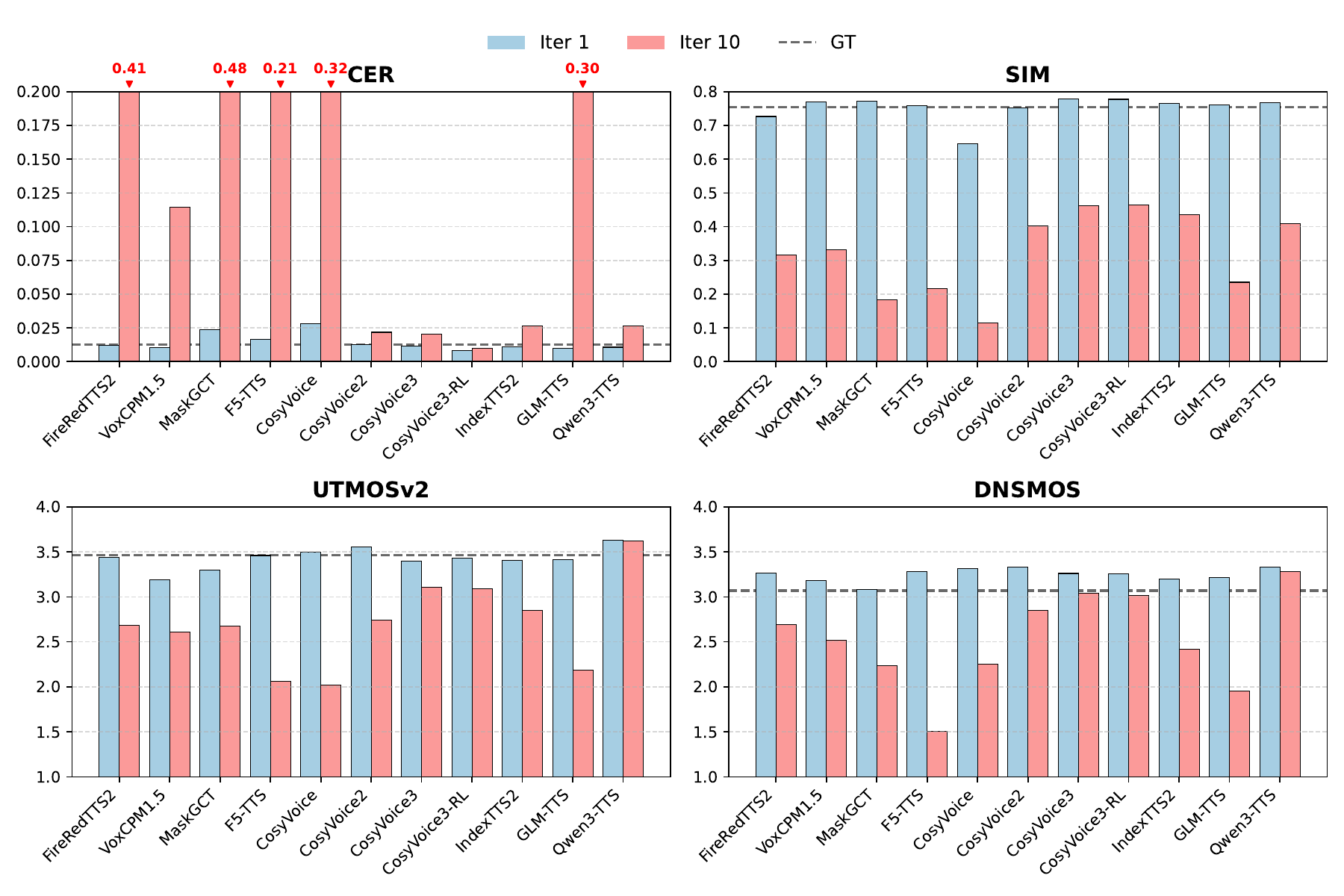}
  \caption{Bar chart of objective metrics for all models on \textbf{Chinese dataset} at the 1st and 10th iterations. The dashed lines indicate the values for real audio.}
  \label{fig:iter1_vs_iter10}
\end{figure*}

\section{Experiments}

\subsection{Evaluated Models}

We evaluate 11 open-source TTS models with strong reported performance. Specifically, we use Qwen3-TTS-12Hz-1.7B-Base for Qwen3-TTS and CosyVoice-300M for CosyVoice. These models cover three major architectural paradigms: autoregressive (AR), non-autoregressive (NAR), and hybrid architectures. Table~\ref{tab:tts_models} summarizes the evaluated systems and their corresponding categories.

\begin{table}[h]
\centering
\caption{Evaluated TTS Models and Architectures}
\label{tab:tts_models}
\begin{tabular}{ll}
\hline
\textbf{Architecture} & \textbf{Model} \\
\hline
\multirow{3}{*}{AR} 
 & FireRedTTS2 \cite{guo2024fireredtts}\\
 & Qwen3-TTS \cite{hu2026qwen3} \\
 & VoxCPM1.5 \cite{zhou2025voxcpm} \\
\hline
\multirow{2}{*}{NAR} 
 & F5-TTS \cite{chen2025f5} \\
 & MaskGCT \cite{wang2024maskgct} \\
\hline
\multirow{6}{*}{Hybrid} 
 & CosyVoice \cite{du2024cosyvoice} \\
 & CosyVoice2 \cite{du2024cosyvoice2} \\
 & CosyVoice3 \cite{du2025cosyvoice} \\
 & CosyVoice3-RL \cite{du2025cosyvoice} \\
 & GLM-TTS \cite{cui2025glm} \\
 & IndexTTS2 \cite{zhou2025indextts2} \\
\hline
\end{tabular}
\end{table}

\subsection{Prompt Configuration}

For the Chinese and emotion datasets, predefined prompt audio lists are provided, and we strictly follow the official evaluation settings. For the English dataset, prior evaluations based on LibriTTS \textit{test-clean} adopt inconsistent configurations, including differences in sample selection and prompt construction, with many implementation details unavailable. To ensure reproducibility and eliminate ambiguity in prompt design, we directly use the ground-truth audio corresponding to the target text as the reference audio for the first synthesis iteration.

\subsection{Evaluation Implementation}

We set max\_iteration to 10 and compute objective metrics for the synthesized speech at each iteration. For the English dataset, all objective metrics are computed using the VERSA toolkit \cite{shi2025versa}. Specifically, WER is calculated with Whisper-large-v3 \cite{radford2023robust}, and SIM is computed using ESPNet \cite{jung2024espnet}. For the Chinese dataset, CER and SIM follow the official evaluation protocol of Seed-TTS-Eval, where Paraformer-zh \cite{gao2022paraformer} is used for speech recognition and a fine-tuned WavLM model \cite{chen2022wavlm} is used for SIM evaluation. DNSMOS and UTMOSv2 are also obtained via the VERSA toolkit. For the emotion dataset, we follow the official CV3-Eval protocol and utilize the emo2vec-large-plus model \cite{ma2024emotion2vec} to report the F1-score for each emotion category, along with the weighted average.

For human evaluation, we include all first-iteration and tenth-iteration synthesized samples from the human-evaluation subset, along with their corresponding real recordings. This results in 2,290 evaluation samples in total. Each sample is independently rated by 5 to 6 annotators, yielding 11,752 annotation records. To prevent loudness discrepancies from biasing perceptual judgments, we apply energy normalization to all audio samples prior to annotation. To ensure data quality, we then perform outlier filtering based on three criteria: inter-annotator consistency, annotation duration, and the discrepancy between subjective and objective scores. This process resulted in the exclusion of approximately 1.2\% of the total annotations.

\section{Results \& Analysis}

\subsection{Low Discriminability and Weak Correlation under Score Saturation}

A fundamental expectation in objective TTS evaluation is that metric scores should be consistent with human judgments and preserve similar system rankings. When analyzing the correlation between conventional evaluation results and human judgments, however, we observe an unexpected and concerning phenomenon: weak correlation between objective metrics and subjective scores across systems.

Fig.~\ref{fig:correlation} presents the Spearman Rank Correlation Coefficients (SRCC) between objective and subjective metrics. Specifically, the utterance-level correlation is calculated based on scores of all individual samples, while the system-level correlation is derived from the overall model rankings. At the first iteration, SIM and CER exhibit weak positive correlations with Speaker Consistency and Content Accuracy at the utterance level. At the system level, however, both metrics maintain relatively strong positive correlations, indicating that they retain some coarse-grained ranking ability across models. In contrast, predictive MOS metrics perform even worse. At both the utterance and system levels, UTMOSv2 and DNSMOS exhibit only very weak correlations with overall naturalness. This implies that, under the prevailing evaluation protocol, these metrics are inadequate for producing reliable system rankings for high-performance TTS models.

To understand the reason behind this phenomenon, we further analyze the distribution of metric scores across systems. As shown in Fig.~\ref{fig:iter1_vs_iter10}, at the first iteration, all evaluated models achieve highly similar scores across the four objective metrics, and some models reach or even surpass the level of ground truth. The standard deviation of SIM (percentage) is 3.83, which decreases to 1.51 after excluding CosyVoice, whose score is substantially lower than those of others. Likewise, the standard deviations of CER (percentage), UTMOSv2, and DNSMOS are only 0.64, 0.12, and 0.07, respectively. These results reveal a pronounced score saturation: performance differences among strong models on these metrics are unfortunately compressed into an extremely and narrow numerical range (within approximately 2\% for all metrics). Under such limited range, the intrinsic bias of evaluation models becomes comparable to, or even larger than, the true performance gap across systems. As a consequence, small fluctuations in metric evaluation can alter system rankings, leading to unreliable results and weak correlation with human evaluation.

\begin{table}[t]
\centering
\caption{\textbf{System-level SRCC} between aggregated metrics and human evaluation on first-iteration synthesized speech. The correlation of objective metrics computed on first-iteration outputs is included as the baseline.}
\label{tab:score_method_correlation_split}
\begin{tabular}{lcc}
\toprule
Score Method 
& SIM
& $1-\mathrm{CER}$ \\
\midrule
Iter1(baseline) 
& 0.6818
& 0.5103 \\

Mean 
& 0.6818
& 0.4829 \\

LWA
& 0.6818
& \textbf{0.5194} \\

EWA
& \textbf{0.7273}
& 0.4282 \\

AUC
& 0.6818
& 0.4282 \\

\midrule
Score Method 
& UTMOSv2 
& DNSMOS \\
\midrule
Iter1(baseline) 
& 0.1182
& 0.0909 \\

Mean
& \textbf{0.4636}
& \textbf{0.2545} \\

LWA
& 0.4364
& 0.2091 \\

EWA
& 0.4273
& 0.1364 \\

AUC
& 0.4545
& \textbf{0.2545} \\

\bottomrule
\end{tabular}
\end{table}

\begin{table*}[t]
\centering
\caption{Objective and subjective evaluation results of different TTS models.
Objective metrics are aggregated using Mean Score (Mean) and \textbf{reported as en/zh}.
Subjective metrics are \textbf{reported at the first and tenth iterations} (iter1 / iter10).
\textbf{Bold} denotes the best result in each sub-column. \uline{Underline} denotes the second-best result in each sub-column.}
\label{tab:tts_overall_results}
\setlength{\tabcolsep}{4.5pt}
\renewcommand{\arraystretch}{1.08}

% 从数据行开始交替着色
\rowcolors{3}{gray!8}{white}

\begin{tabular}{lccc|ccc}
\toprule
\rowcolor{white}
\multirow{2}{*}{\textbf{TTS Models}}
& \multicolumn{3}{c}{\textbf{Objective Metrics (en/zh)}}
& \multicolumn{3}{c}{\textbf{Subjective Metrics (iter1 / iter10)}} \\
\rowcolor{white}
\cmidrule(lr){2-4} \cmidrule(lr){5-7}
& WER / CER$\downarrow$ & SIM$\uparrow$ & UTMOSv2$\uparrow$
& Content Acc.$\uparrow$ & Spk. Consistency$\uparrow$ & Naturalness$\uparrow$ \\
\midrule
    CosyVoice
      &  8.88 / 11.18 & 19.34 / 23.70 & 2.98 / 2.76
      &  4.75 /  3.25 &  3.48 /  1.01 & 3.93 / 1.60 \\
    CosyVoice2
      &  2.77 /  1.51 & 61.34 / 56.88 & 3.80 / 3.22
      &  4.84 /  4.62 &  4.45 /  1.53 & 4.08 / 2.95 \\
    CosyVoice3
      &  2.36 /  \uline{1.44} & \uline{66.88} / \uline{62.14} & 3.82 / \uline{3.33}
      &  4.80 /  4.64 & \textbf{4.55} /  \uline{1.90} & 4.04 / \uline{3.18} \\
    CosyVoice3-RL
      &  2.27 / \textbf{0.90} & \textbf{67.32} / \textbf{62.17} & \uline{3.83} / \uline{3.33}
      &  4.84 / \textbf{4.73} &  \uline{4.53} / \textbf{1.98} & 4.07 / 3.16 \\
    F5-TTS
      & 17.40 /  5.33 & 48.74 / 49.23 & 3.09 / 2.92
      &  4.79 /  2.78 &  4.49 /  1.05 & 4.10 / 1.43 \\
    FireRedTTS2
      & 32.35 / 18.33 & 32.28 / 50.05 & 3.18 / 3.10
      &  4.84 /  3.28 &  4.35 /  1.26 & 3.95 / 2.01 \\
    GLM-TTS
      &  5.35 /  8.53 & 45.38 / 49.44 & 3.20 / 2.74
      &  4.84 /  2.64 &  4.49 /  1.07 & 4.07 / 1.44 \\
    IndexTTS2
      & \textbf{2.22} /  1.72 & 62.06 / 58.50 & 3.46 / 3.22
      & \textbf{4.88} /  4.59 &  4.49 /  1.40 & \textbf{4.30} / 2.56 \\
    MaskGCT
      & 11.88 / 17.26 & 47.21 / 43.89 & 3.29 / 2.95
      &  4.78 /  2.42 &  4.43 /  1.02 & 3.81 / 1.31 \\
    Qwen3-TTS
      &  \uline{2.26} /  1.68 & 64.73 / 57.19 & \textbf{4.04} / \textbf{3.68}
      &  \uline{4.86} /  \uline{4.69} &  4.52 /  1.82 & \uline{4.27} / \textbf{3.79} \\
    VoxCPM1.5
      &  9.15 /  4.86 & 52.70 / 53.36 & 3.12 / 2.93
      &  4.85 /  4.22 &  4.47 /  1.34 & 4.11 / 2.59 \\
\bottomrule
\end{tabular}
\end{table*}

\subsection{Restoring Discriminability and Human Correlation via Difference Amplification}

To address the score saturation observed under the conventional protocol, we apply the proposed iterative evaluation strategy and re-examine model performance across multiple synthesis rounds. This strategy preserves the use of existing objective metrics while enhancing their discriminative capacity in the strong-model regime.

When evaluating speech generated at the 10th iteration, performance differences among models become substantially more pronounced. The standard deviations of SIM, CER, UTMOSv2, and DNSMOS increase to 12.15, 17.62, 0.48, and 0.52, respectively, compared with the highly concentrated distributions observed at the first iteration. This underscores the efficacy of iterative synthesis in magnifying performance disparities across models. For example, on the SIM metric, the best-performing model, CosyVoice3-RL, achieves a score of 46.34, while weaker models remain around 20, and the lowest-performing model, CosyVoice, drops to 11.41. A similar trend occurs for CER. Although all models perform well at the first iteration (with the worst result being 2.83), after 10 iterations, several models exhibit CER values exceeding 10. Notably, FireRedTTS2, and MaskGCT exceed a WER of 40, suggesting a clear degradation in semantic consistency after iterative synthesis. Consistent trends are also observed for UTMOSv2 and DNSMOS. While all models demonstrate comparable speech quality at the first iteration, iterative synthesis progressively exposes robustness differences across systems, reflected in degradations such as background noise, artifacts, unstable prosody, and pitch distortion in almost all models except Qwen3-TTS.

More importantly, iterative evaluation substantially improves the alignment between objective metrics and human subjective judgments. After ten iterations, the correlation between objective metrics and human evaluations increases significantly. As shown in Fig.~\ref{fig:correlation}, at the utterance level, the SRCC between each metric and its corresponding subjective dimension exceed 0.6, indicating a moderate correlation. At the system level, the SRCC values for all objective metrics exceed 0.8, demonstrating strong consistency with human rankings. This improvement arises because iterative synthesis amplifies inter-model differences, mitigating intrinsic errors in the evaluation models.

\begin{figure}[t]
  \centering
  \includegraphics[width=\columnwidth]{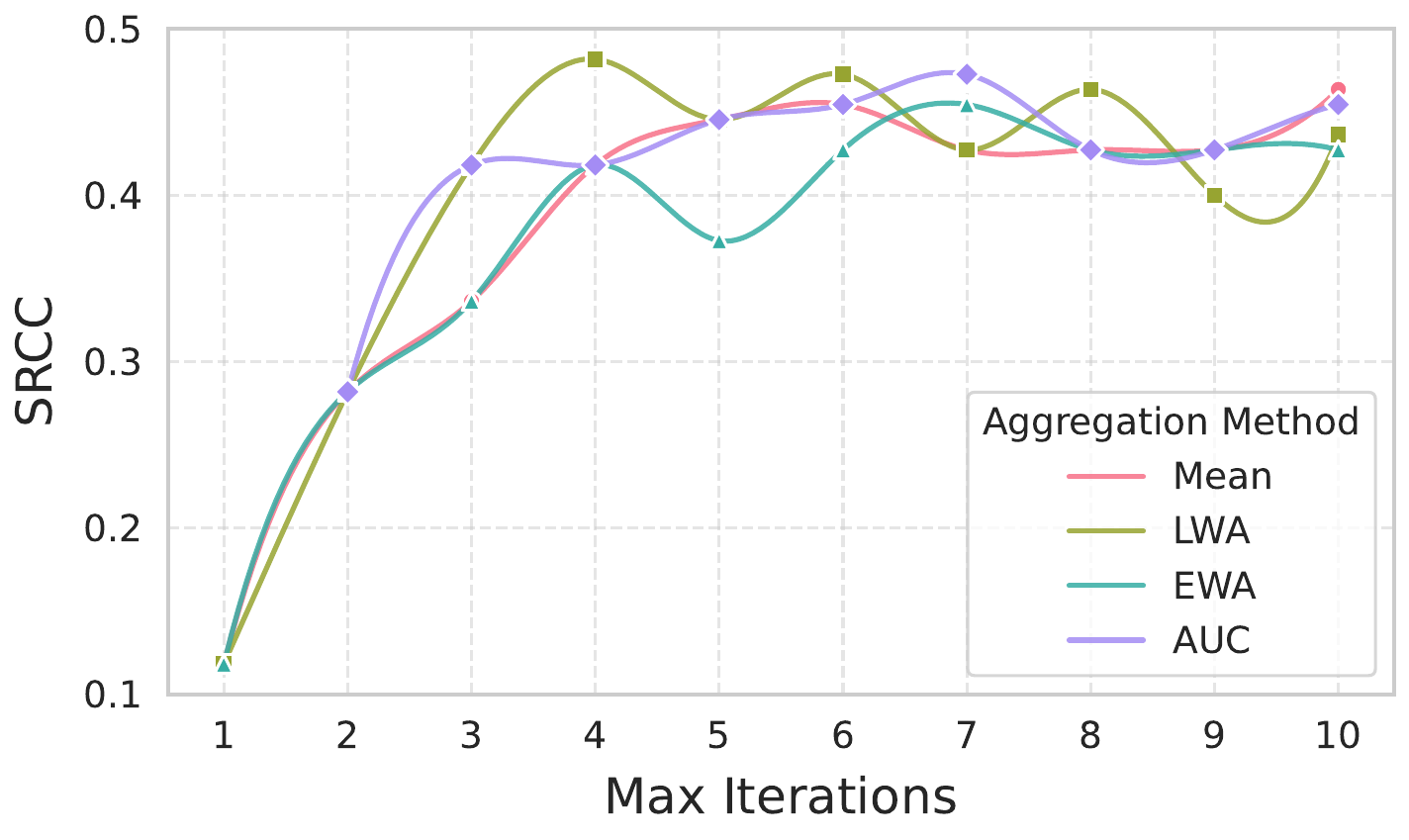}
  \caption{System-level SRCC between aggregated UTMOSv2 scores and human ratings of first-iteration speech, computed under different maximum iteration numbers.}
  \label{fig:srcc_curve}
\end{figure}

In practical applications, users are primarily concerned with first-iteration speech quality. Therefore, we further examine whether aggregated iterative scores correlate with human judgments of first-iteration outputs. As shown in Table~\ref{tab:score_method_correlation_split}, we compute the system-level SRCC between model rankings from aggregated objective scores and those from human ratings on first-iteration speech. Most strategies perform on par with or slightly outperform the first-iteration baseline on SIM and 1-CER. In contrast, UTMOSv2 and DNSMOS show clear improvements after aggregation. For UTMOSv2, the correlation increases substantially from 0.1182 to 0.4636 under the Mean strategy. DNSMOS exhibits a similar trend, improving from 0.0909 to 0.2545 under both Mean and AUC. These results indicate that aggregation notably enhances the system-level consistency of predictive MOS metrics with human judgments of first-iteration speech. We further investigate the impact of different maximum iteration numbers used for metric aggregation on the system-level SRCC with human ratings on first-iteration. As shown in Fig.~\ref{fig:srcc_curve}, the SRCC reaches a comparable level to that obtained with 10 iterations when the maximum iteration number is around 5. Therefore, under limited computational resources, we recommend using 5 as the maximum iteration number to balance computational cost and human alignment.

\subsection{Report of TTS Model Capability}

\begin{table}[t]
\centering
\caption{Emotion classification F1-score (\%) aggregated using the Mean Score strategy on the emotion dataset for different TTS models.}
\label{tab:emotion_results}
\begin{tabular}{lcccc}
\toprule
\textbf{TTS Models} & \textbf{Angry} & \textbf{Happy} & \textbf{Sad} & \textbf{Weighted Avg} \\
\midrule
GT            & 75.10 & 85.70 & 78.20 & 79.70 \\
\midrule
CosyVoice     & 30.85 & 30.21 & \textbf{57.29} & 39.57 \\
CosyVoice2    & 50.69 & 54.93 & 47.19 & 50.94 \\
CosyVoice3    & 59.69 & 60.91 & 46.30 & 55.65 \\
CosyVoice3-RL & 63.87 & 53.51 & 34.00 & 50.47 \\
F5-TTS        & 52.47 & 56.14 & 34.88 & 47.83 \\
FireRedTTS2   & 38.87 & 54.43 & 50.69 & 47.88 \\
GLM-TTS       & 47.77 & 54.35 & 55.57 & 52.55 \\
IndexTTS2    & \textbf{73.26} & \textbf{71.42} & 47.92 & \textbf{64.69} \\
MaskGCT       & 38.10 & 49.72 & \uline{55.87} & 47.89 \\
Qwen3-TTS     & 30.77 & \uline{62.52} & 31.60 & 41.58 \\
VoxCPM1.5        & \uline{64.33} & 56.77 & 52.34 & \uline{57.82} \\
\bottomrule
\end{tabular}
\end{table}

We report the evaluation results of 11 TTS models on three datasets. Objective metrics (WER/CER, SIM, and UTMOSv2) are aggregated using the Mean Score strategy. Subjective metrics are reported at the first and tenth iterations on the human-evaluation subset, covering Content Accuracy, Speaker Consistency, and Overall Naturalness. Emotion cloning performance is evaluated using aggregated F1-score on the emotion dataset.

As shown in Table~\ref{tab:tts_overall_results}, CosyVoice3, CosyVoice3-RL, CosyVoice2, IndexTTS2, and Qwen3-TTS demonstrate consistently strong performance across multiple dimensions. Among them, CosyVoice3-RL achieves the best CER (0.90) and the highest SIM scores (67.32 / 62.17). CosyVoice3 attains comparable WER/CER (2.36 / 1.44) and SIM (66.88 / 62.14), indicating similar overall performance. IndexTTS2 achieves the lowest WER (2.22) and obtains the highest first-iteration subjective scores in Content Accuracy (4.88) and Naturalness (4.30), indicating particularly strong performance in initial synthesis quality. Qwen3-TTS achieves the highest UTMOSv2 scores (4.04 / 3.68), with a clear margin over the second-ranked CosyVoice3. Moreover, its tenth-iteration Naturalness score (3.79) is the highest among all models, reflecting comparatively stable perceptual quality under iterative evaluation.

In contrast, certain models exhibit clear degradation patterns during iterative synthesis. FireRedTTS2 reports substantially higher WER/CER than other systems. Examination of its synthesized samples shows that, during iteration, it may generate extremely short utterances, excessively long silent segments, or even audio content unrelated to the target text. When such outputs are reused as reference inputs in subsequent iterations, the mismatch between speech and text is further amplified, leading to markedly elevated WER/CER. This behavior is directly reflected in the objective metrics and indicates limited stability under iterative conditions. F5-TTS demonstrates a gradual increase in speaking rate and increasingly monotonous intonation as iteration proceeds, which negatively affects its performance in both objective and subjective evaluations. CosyVoice exhibits noticeable electrical noise during iterative synthesis, with speech becoming progressively muffled and distorted. Similarly, GLM-TTS and MaskGCT show reduced speech clarity accompanied by unnatural and irregular intonation patterns. These phenomena are consistent with their declines in evaluation metrics under iterative settings.

Emotion cloning results are summarized in Table~\ref{tab:emotion_results}. IndexTTS2 achieves the highest weighted average F1-score (64.69) and ranks first in both Angry (73.26) and Happy (71.42), indicating strong emotion controllability. CosyVoice3 and VoxCPM1.5 also demonstrate competitive overall emotion performance (55.65 and 57.82). Although CosyVoice achieves the highest F1-score in the Sad category (57.29), this result is attributable to a systematic bias in its generation behavior. As iteration increases, CosyVoice outputs tend to converge toward a Sad-like emotional tone, leading to high recall for that category while substantially degrading performance in others. This behavior does not reflect genuine emotion modeling capability, but rather a tendency toward homogenized emotional outputs.

\subsection{Cross-Model Iteration Analysis}

\begin{figure}[t]
  \centering
  \includegraphics[width=\columnwidth]{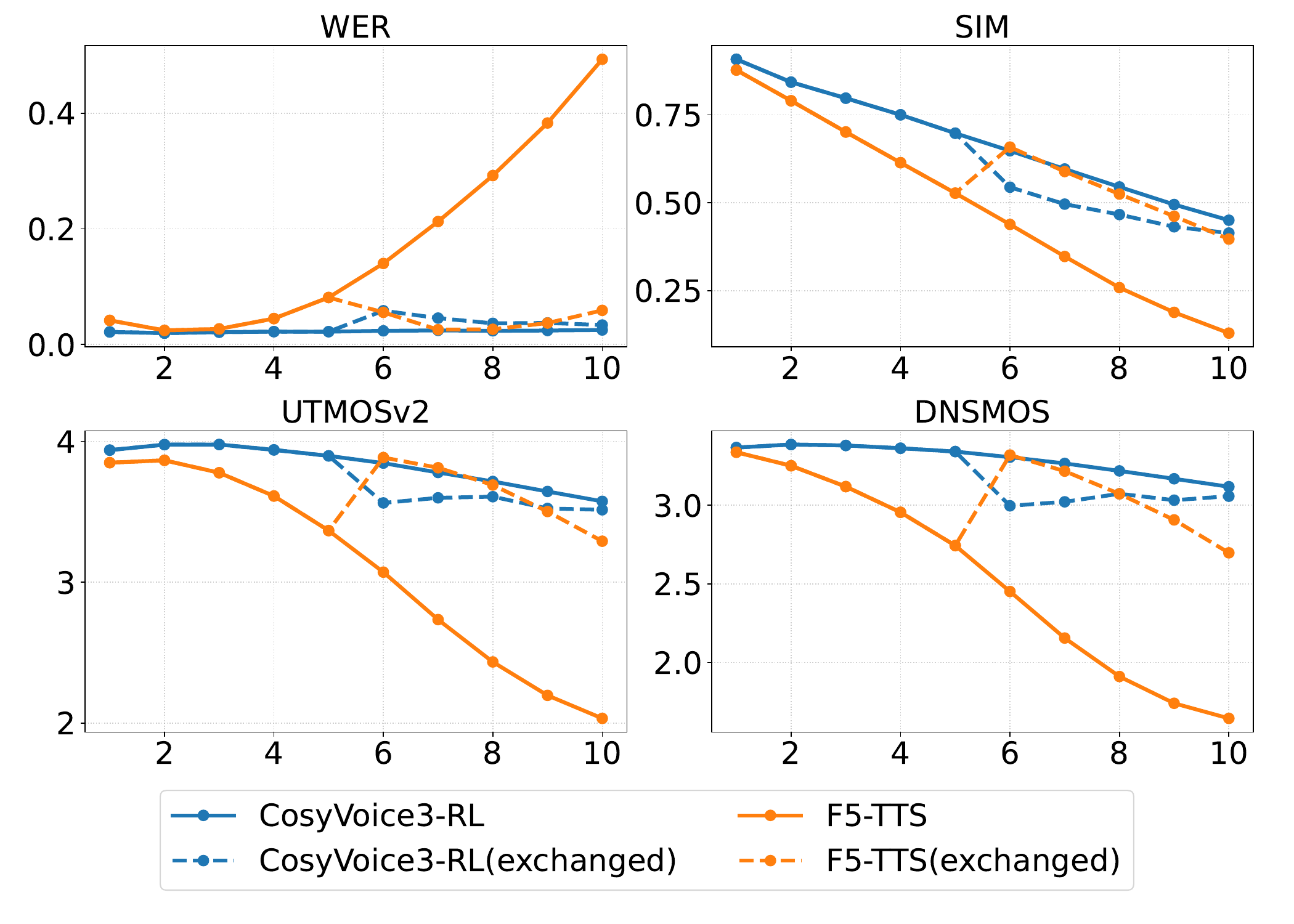}
  \caption{Cross-model iterative evaluation curves of CosyVoice3-RL and F5-TTS. Solid lines denote the original trajectories without reference swapping, while dashed lines represent the trajectories after exchanging reference audio at the 6th iteration.}
  \label{fig:cross-model}
\end{figure}

The quality degradation observed during iterative synthesis may arise from two potential factors: (1) progressive deterioration of reference audio quality, which weakens its effectiveness as a conditioning signal; (2) distribution shift, where iteratively generated reference audio gradually deviates from the model’s training distribution, thereby impairing generalization. To disentangle these two factors, we conduct a cross-model iteration experiment. We select CosyVoice3-RL, which demonstrates strong robustness, and F5-TTS, which exhibits rapid degradation under iteration. At the 6th iteration, we exchange their reference audio while keeping all other configurations unchanged, and then continue the iterative process.

The results are illustrated in Fig.~\ref{fig:cross-model}. When F5-TTS is conditioned on reference audio generated by CosyVoice3-RL, its performance increases sharply at the exchange point across all metrics, approaching the original trajectory of CosyVoice3-RL. This observation suggests that the provision of high-quality reference audio yields immediate gains in synthesis performance. However, in subsequent iterations, F5-TTS again experiences degradation and diverges from CosyVoice3-RL, suggesting that it lacks the intrinsic robustness required to sustain quality under repeated iteration. Conversely, when CosyVoice3-RL is conditioned on reference audio generated by F5-TTS, its performance drops at the exchange point, reflecting the lower quality of the injected reference. Nevertheless, as iteration proceeds, the performance gap between the exchanged trajectory and the original trajectory gradually narrows. 

These observations indicate that the observed performance decay in iterative synthesis is primarily driven by progressive reference quality degradation, rather than catastrophic out-of-distribution effects. Furthermore, the experiment highlights clear differences in model robustness: stronger models can maintain stable generation behavior, whereas weaker models remain highly sensitive to reference quality and consequently tend to collapse under continued iteration.

\section{Limitations}

While improving discriminative power and human alignment, the I2D framework entails higher computational costs due to repeated synthesis and metric computation. Its iterative nature also tends to favor model stability over expressive diversity; we therefore recommend augmenting this approach with diversity-oriented measures for a more holistic assessment. Furthermore, the evaluation of naturalness is not only influenced by a model's intrinsic capability but is also heavily conditioned by the reference audio. In zero-shot settings, this leads to a certain conflict between naturalness and speaker similarity, particularly when the reference audio is of suboptimal quality. Finally, our study focuses on open-source TTS systems, as commercial models remain inaccessible due to cost and access constraints.

\section{Conclusion}

We propose an iterative evaluation framework (I2D) for TTS that aims to mitigate the low discriminability and weak human correlation caused by score saturation under conventional evaluation. By repeatedly synthesizing speech and aggregating metric scores, the proposed method amplifies inter-model performance differences, reduces the relative impact of intrinsic evaluation bias, and restores ranking reliability among strong TTS systems. The framework further provides insight into model behavior under multiple iterations, revealing stability differences that are not observable in standard evaluation settings. Overall, the I2D framework offers a practical and scalable approach toward more reliable and discriminative automated TTS assessment.

\section{Generative AI Use Disclosure}
We acknowledge the use of AI tools for pure language polishing in preparing of this manuscript. The authors verified the generated text and take full responsibility for the content.

\section{Acknowledgments}
This work utilized seed-tts-eval and CV3-Eval (sources: https://github.com/BytedanceSpeech/seed-tts-eval, https://github.com/FunAudioLLM/CV3-Eval). The authors confirm that the use of the above datasets and code in this paper is strictly for academic research purposes and not for any commercial activities.

\bibliographystyle{IEEEtran}
\bibliography{mybib}

@article{anastassiou2024seed,
  title={Seed-tts: A family of high-quality versatile speech generation models},
  author={Anastassiou, Philip and Chen, Jiawei and Chen, Jitong and Chen, Yuanzhe and Chen, Zhuo and Chen, Ziyi and Cong, Jian and Deng, Lelai and Ding, Chuang and Gao, Lu and others},
  journal={arXiv preprint arXiv:2406.02430},
  year={2024}
}

@article{du2025cosyvoice,
  title={Cosyvoice 3: Towards in-the-wild speech generation via scaling-up and post-training},
  author={Du, Zhihao and Gao, Changfeng and Wang, Yuxuan and Yu, Fan and Zhao, Tianyu and Wang, Hao and Lv, Xiang and Wang, Hui and Ni, Chongjia and Shi, Xian and others},
  journal={arXiv preprint arXiv:2505.17589},
  year={2025}
}

@article{zhou2025indextts2,
  title={IndexTTS2: A Breakthrough in Emotionally Expressive and Duration-Controlled Auto-Regressive Zero-Shot Text-to-Speech},
  author={Zhou, Siyi and Zhou, Yiquan and He, Yi and Zhou, Xun and Wang, Jinchao and Deng, Wei and Shu, Jingchen},
  journal={arXiv preprint arXiv:2506.21619},
  year={2025}
}

@article{hu2026qwen3,
  title={Qwen3-TTS Technical Report},
  author={Hu, Hangrui and Zhu, Xinfa and He, Ting and Guo, Dake and Zhang, Bin and Wang, Xiong and Guo, Zhifang and Jiang, Ziyue and Hao, Hongkun and Guo, Zishan and others},
  journal={arXiv preprint arXiv:2601.15621},
  year={2026}
}

@article{yang2025towards,
  title={Towards Responsible Evaluation for Text-to-Speech},
  author={Yang, Yifan and Wang, Hui and Han, Bing and Liu, Shujie and Li, Jinyu and Qin, Yong and Chen, Xie},
  journal={arXiv preprint arXiv:2510.06927},
  year={2025}
}

@article{tee2025sp,
  title={SP-MCQA: Evaluating Intelligibility of TTS Beyond the Word Level},
  author={Tee, Hitomi Jin Ling and Wang, Chaoren and Zhang, Zijie and Wu, Zhizheng},
  journal={arXiv preprint arXiv:2510.26190},
  year={2025}
}

@article{chiang2023we,
  title={Why we should report the details in subjective evaluation of TTS more rigorously},
  author={Chiang, Cheng-Han and Huang, Wei-Ping and Lee, Hung-yi},
  journal={arXiv preprint arXiv:2306.02044},
  year={2023}
}

@article{lo2019mosnet,
  title={Mosnet: Deep learning based objective assessment for voice conversion},
  author={Lo, Chen-Chou and Fu, Szu-Wei and Huang, Wen-Chin and Wang, Xin and Yamagishi, Junichi and Tsao, Yu and Wang, Hsin-Min},
  journal={arXiv preprint arXiv:1904.08352},
  year={2019}
}

@inproceedings{reddy2021dnsmos,
  title={DNSMOS: A non-intrusive perceptual objective speech quality metric to evaluate noise suppressors},
  author={Reddy, Chandan KA and Gopal, Vishak and Cutler, Ross},
  booktitle={ICASSP 2021-2021 IEEE International Conference on Acoustics, Speech and Signal Processing (ICASSP)},
  pages={6493--6497},
  year={2021},
  organization={IEEE}
}

@article{saeki2022utmos,
  title={Utmos: Utokyo-sarulab system for voicemos challenge 2022},
  author={Saeki, Takaaki and Xin, Detai and Nakata, Wataru and Koriyama, Tomoki and Takamichi, Shinnosuke and Saruwatari, Hiroshi},
  journal={arXiv preprint arXiv:2204.02152},
  year={2022}
}

@article{wang2026urgentmos,
  title={UrgentMOS: Unified Multi-Metric and Preference Learning for Robust Speech Quality Assessment},
  author={Wang, Wei and Zhang, Wangyou and Li, Chenda and Wang, Jiahe and Cornell, Samuele and Sach, Marvin and Saijo, Kohei and Fu, Yihui and Ni, Zhaoheng and Han, Bing and others},
  journal={arXiv preprint arXiv:2601.18438},
  year={2026}
}

@inproceedings{huang2024voicemos,
  title={The VoiceMOS challenge 2024: Beyond speech quality prediction},
  author={Huang, Wen-Chin and Fu, Szu-Wei and Cooper, Erica and Zezario, Ryandhimas E and Toda, Tomoki and Wang, Hsin-Min and Yamagishi, Junichi and Tsao, Yu},
  booktitle={2024 IEEE Spoken Language Technology Workshop (SLT)},
  pages={803--810},
  year={2024},
  organization={IEEE}
}

@article{chen2024takin,
  title={Takin: A cohort of superior quality zero-shot speech generation models},
  author={Chen, Sijing and Feng, Yuan and He, Laipeng and He, Tianwei and He, Wendi and Hu, Yanni and Lin, Bin and Lin, Yiting and Pan, Yu and Tan, Pengfei and others},
  journal={arXiv preprint arXiv:2409.12139},
  year={2024}
}

@article{wang2025spark,
  title={Spark-tts: An efficient llm-based text-to-speech model with single-stream decoupled speech tokens},
  author={Wang, Xinsheng and Jiang, Mingqi and Ma, Ziyang and Zhang, Ziyu and Liu, Songxiang and Li, Linqin and Liang, Zheng and Zheng, Qixi and Wang, Rui and Feng, Xiaoqin and others},
  journal={arXiv preprint arXiv:2503.01710},
  year={2025}
}

@article{xie2025fireredtts,
  title={Fireredtts-2: Towards long conversational speech generation for podcast and chatbot},
  author={Xie, Kun and Shen, Feiyu and Li, Junjie and Xie, Fenglong and Tang, Xu and Hu, Yao},
  journal={arXiv preprint arXiv:2509.02020},
  year={2025}
}

@article{liao2024fish,
  title={Fish-speech: Leveraging large language models for advanced multilingual text-to-speech synthesis},
  author={Liao, Shijia and Wang, Yuxuan and Li, Tianyu and Cheng, Yifan and Zhang, Ruoyi and Zhou, Rongzhi and Xing, Yijin},
  journal={arXiv preprint arXiv:2411.01156},
  year={2024}
}

@article{casanova2024xtts,
  title={Xtts: a massively multilingual zero-shot text-to-speech model},
  author={Casanova, Edresson and Davis, Kelly and G{\"o}lge, Eren and G{\"o}knar, G{\"o}rkem and Gulea, Iulian and Hart, Logan and Aljafari, Aya and Meyer, Joshua and Morais, Reuben and Olayemi, Samuel and others},
  journal={arXiv preprint arXiv:2406.04904},
  year={2024}
}

@inproceedings{gao2023e3,
  title={E3 tts: Easy end-to-end diffusion-based text to speech},
  author={Gao, Yuan and Morioka, Nobuyuki and Zhang, Yu and Chen, Nanxin},
  booktitle={2023 IEEE Automatic Speech Recognition and Understanding Workshop (ASRU)},
  pages={1--8},
  year={2023},
  organization={IEEE}
}

@article{ju2024naturalspeech,
  title={Naturalspeech 3: Zero-shot speech synthesis with factorized codec and diffusion models},
  author={Ju, Zeqian and Wang, Yuancheng and Shen, Kai and Tan, Xu and Xin, Detai and Yang, Dongchao and Liu, Yanqing and Leng, Yichong and Song, Kaitao and Tang, Siliang and others},
  journal={arXiv preprint arXiv:2403.03100},
  year={2024}
}

@article{wang2024maskgct,
  title={Maskgct: Zero-shot text-to-speech with masked generative codec transformer},
  author={Wang, Yuancheng and Zhan, Haoyue and Liu, Liwei and Zeng, Ruihong and Guo, Haotian and Zheng, Jiachen and Zhang, Qiang and Zhang, Xueyao and Zhang, Shunsi and Wu, Zhizheng},
  journal={arXiv preprint arXiv:2409.00750},
  year={2024}
}

@inproceedings{chen2025f5,
  title={F5-tts: A fairytaler that fakes fluent and faithful speech with flow matching},
  author={Chen, Yushen and Niu, Zhikang and Ma, Ziyang and Deng, Keqi and Wang, Chunhui and JianZhao, JianZhao and Yu, Kai and Chen, Xie},
  booktitle={Proceedings of the 63rd Annual Meeting of the Association for Computational Linguistics (Volume 1: Long Papers)},
  pages={6255--6271},
  year={2025}
}

@article{guo2024fireredtts,
  title={Fireredtts: A foundation text-to-speech framework for industry-level generative speech applications},
  author={Guo, Hao-Han and Hu, Yao and Liu, Kun and Shen, Fei-Yu and Tang, Xu and Wu, Yi-Chen and Xie, Feng-Long and Xie, Kun and Xu, Kai-Tuo},
  journal={arXiv preprint arXiv:2409.03283},
  year={2024}
}

@article{huang2022voicemos,
  title={The voicemos challenge 2022},
  author={Huang, Wen-Chin and Cooper, Erica and Tsao, Yu and Wang, Hsin-Min and Toda, Tomoki and Yamagishi, Junichi},
  journal={arXiv preprint arXiv:2203.11389},
  year={2022}
}

@article{manakul2025audiojudge,
  title={Audiojudge: Understanding what works in large audio model based speech evaluation},
  author={Manakul, Potsawee and Gan, Woody Haosheng and Ryan, Michael J and Khan, Ali Sartaz and Sirichotedumrong, Warit and Pipatanakul, Kunat and Held, William and Yang, Diyi},
  journal={arXiv preprint arXiv:2507.12705},
  year={2025}
}

@article{zhang2025speechjudge,
  title={SpeechJudge: Towards Human-Level Judgment for Speech Naturalness},
  author={Zhang, Xueyao and Wang, Chaoren and Liao, Huan and Li, Ziniu and Wang, Yuancheng and Wang, Li and Jia, Dongya and Chen, Yuanzhe and Li, Xiulin and Chen, Zhuo and others},
  journal={arXiv preprint arXiv:2511.07931},
  year={2025}
}

@article{wang2025speechllm,
  title={SpeechLLM-as-Judges: Towards General and Interpretable Speech Quality Evaluation},
  author={Wang, Hui and Zhao, Jinghua and Yang, Yifan and Liu, Shujie and Chen, Junyang and Zhang, Yanzhe and Zhao, Shiwan and Li, Jinyu and Zhou, Jiaming and Sun, Haoqin and others},
  journal={arXiv preprint arXiv:2510.14664},
  year={2025}
}

@article{xu2025qwen2,
  title={Qwen2. 5-omni technical report},
  author={Xu, Jin and Guo, Zhifang and He, Jinzheng and Hu, Hangrui and He, Ting and Bai, Shuai and Chen, Keqin and Wang, Jialin and Fan, Yang and Dang, Kai and others},
  journal={arXiv preprint arXiv:2503.20215},
  year={2025}
}

@article{zheng2023judging,
  title={Judging llm-as-a-judge with mt-bench and chatbot arena},
  author={Zheng, Lianmin and Chiang, Wei-Lin and Sheng, Ying and Zhuang, Siyuan and Wu, Zhanghao and Zhuang, Yonghao and Lin, Zi and Li, Zhuohan and Li, Dacheng and Xing, Eric and others},
  journal={Advances in neural information processing systems},
  volume={36},
  pages={46595--46623},
  year={2023}
}

@article{zen2019libritts,
  title={Libritts: A corpus derived from librispeech for text-to-speech},
  author={Zen, Heiga and Dang, Viet and Clark, Rob and Zhang, Yu and Weiss, Ron J and Jia, Ye and Chen, Zhifeng and Wu, Yonghui},
  journal={arXiv preprint arXiv:1904.02882},
  year={2019}
}

@inproceedings{guo2021didispeech,
  title={Didispeech: A large scale mandarin speech corpus},
  author={Guo, Tingwei and Wen, Cheng and Jiang, Dongwei and Luo, Ne and Zhang, Ruixiong and Zhao, Shuaijiang and Li, Wubo and Gong, Cheng and Zou, Wei and Han, Kun and others},
  booktitle={ICASSP 2021-2021 IEEE International Conference on Acoustics, Speech and Signal Processing (ICASSP)},
  pages={6968--6972},
  year={2021},
  organization={IEEE}
}

@article{ma2024emobox,
  title={Emobox: Multilingual multi-corpus speech emotion recognition toolkit and benchmark},
  author={Ma, Ziyang and Chen, Mingjie and Zhang, Hezhao and Zheng, Zhisheng and Chen, Wenxi and Li, Xiquan and Ye, Jiaxin and Chen, Xie and Hain, Thomas},
  journal={arXiv preprint arXiv:2406.07162},
  year={2024}
}

@inproceedings{xu2024secap,
  title={Secap: Speech emotion captioning with large language model},
  author={Xu, Yaoxun and Chen, Hangting and Yu, Jianwei and Huang, Qiaochu and Wu, Zhiyong and Zhang, Shi-Xiong and Li, Guangzhi and Luo, Yi and Gu, Rongzhi},
  booktitle={Proceedings of the AAAI Conference on Artificial Intelligence},
  volume={38},
  number={17},
  pages={19323--19331},
  year={2024}
}

@inproceedings{baba2024t05,
  title={The t05 system for the voicemos challenge 2024: Transfer learning from deep image classifier to naturalness mos prediction of high-quality synthetic speech},
  author={Baba, Kaito and Nakata, Wataru and Saito, Yuki and Saruwatari, Hiroshi},
  booktitle={2024 IEEE Spoken Language Technology Workshop (SLT)},
  pages={818--824},
  year={2024},
  organization={IEEE}
}

@inproceedings{shi2025versa,
  title={VERSA: A versatile evaluation toolkit for speech, audio, and music},
  author={Shi, Jiatong and Shim, Hye-jin and Tian, Jinchuan and Arora, Siddhant and Wu, Haibin and Petermann, Darius and Yip, Jia Qi and Zhang, You and Tang, Yuxun and Zhang, Wangyou and others},
  booktitle={Proceedings of the 2025 Conference of the Nations of the Americas Chapter of the Association for Computational Linguistics: Human Language Technologies (System Demonstrations)},
  pages={191--209},
  year={2025}
}

@inproceedings{radford2023robust,
  title={Robust speech recognition via large-scale weak supervision},
  author={Radford, Alec and Kim, Jong Wook and Xu, Tao and Brockman, Greg and McLeavey, Christine and Sutskever, Ilya},
  booktitle={International conference on machine learning},
  pages={28492--28518},
  year={2023},
  organization={PMLR}
}

@article{jung2024espnet,
  title={Espnet-spk: full pipeline speaker embedding toolkit with reproducible recipes, self-supervised front-ends, and off-the-shelf models},
  author={Jung, Jee-weon and Zhang, Wangyou and Shi, Jiatong and Aldeneh, Zakaria and Higuchi, Takuya and Theobald, Barry-John and Abdelaziz, Ahmed Hussen and Watanabe, Shinji},
  journal={arXiv preprint arXiv:2401.17230},
  year={2024}
}

@article{gao2022paraformer,
  title={Paraformer: Fast and accurate parallel transformer for non-autoregressive end-to-end speech recognition},
  author={Gao, Zhifu and Zhang, Shiliang and McLoughlin, Ian and Yan, Zhijie},
  journal={arXiv preprint arXiv:2206.08317},
  year={2022}
}

@article{chen2022wavlm,
  title={Wavlm: Large-scale self-supervised pre-training for full stack speech processing},
  author={Chen, Sanyuan and Wang, Chengyi and Chen, Zhengyang and Wu, Yu and Liu, Shujie and Chen, Zhuo and Li, Jinyu and Kanda, Naoyuki and Yoshioka, Takuya and Xiao, Xiong and others},
  journal={IEEE Journal of Selected Topics in Signal Processing},
  volume={16},
  number={6},
  pages={1505--1518},
  year={2022},
  publisher={IEEE}
}

@inproceedings{ma2024emotion2vec,
  title={emotion2vec: Self-supervised pre-training for speech emotion representation},
  author={Ma, Ziyang and Zheng, Zhisheng and Ye, Jiaxin and Li, Jinchao and Gao, Zhifu and Zhang, Shiliang and Chen, Xie},
  booktitle={Findings of the Association for Computational Linguistics: ACL 2024},
  pages={15747--15760},
  year={2024}
}

@article{zhou2025voxcpm,
  title={Voxcpm: Tokenizer-free TTS for context-aware speech generation and true-to-life voice cloning},
  author={Zhou, Yixuan and Zeng, Guoyang and Liu, Xin and Li, Xiang and Yu, Renjie and Wang, Ziyang and Ye, Runchuan and Sun, Weiyue and Gui, Jiancheng and Li, Kehan and others},
  journal={arXiv preprint arXiv:2509.24650},
  year={2025}
}

@article{du2024cosyvoice,
  title={Cosyvoice: A scalable multilingual zero-shot text-to-speech synthesizer based on supervised semantic tokens},
  author={Du, Zhihao and Chen, Qian and Zhang, Shiliang and Hu, Kai and Lu, Heng and Yang, Yexin and Hu, Hangrui and Zheng, Siqi and Gu, Yue and Ma, Ziyang and others},
  journal={arXiv preprint arXiv:2407.05407},
  year={2024}
}

@article{du2024cosyvoice2,
  title={Cosyvoice 2: Scalable streaming speech synthesis with large language models},
  author={Du, Zhihao and Wang, Yuxuan and Chen, Qian and Shi, Xian and Lv, Xiang and Zhao, Tianyu and Gao, Zhifu and Yang, Yexin and Gao, Changfeng and Wang, Hui and others},
  journal={arXiv preprint arXiv:2412.10117},
  year={2024}
}

@article{cui2025glm,
  title={Glm-tts technical report},
  author={Cui, Jiayan and Yang, Zhihan and Li, Naihan and Tian, Jiankun and Ma, Xingyu and Zhang, Yi and Chen, Guangyu and Yang, Runxuan and Cheng, Yuqing and Zhou, Yizhi and others},
  journal={arXiv preprint arXiv:2512.14291},
  year={2025}
}

\end{document}